\documentclass[english,twocolumn,aps,prl,showpacs]{revtex4-1}

\usepackage[T1]{fontenc}
\usepackage[latin1]{inputenc}
\usepackage{graphicx}
\usepackage{amssymb}
\usepackage{amsfonts}
\usepackage{amsmath}

\makeatletter

\DeclareMathAlphabet{\mathsfsl}{OT1}{cmss}{m}{sl}

\input{epsf}

\makeatother

\usepackage{babel}
\makeatother

\begin{document}

\title{Dynamic spin response of a strongly interacting Fermi gas}

\author{S. Hoinka$^1$, M. Lingham$^1$, M. Delehaye$^{1,2}$, and C. J. Vale$^1$}
\affiliation{$^1$Centre for Atom Optics and Ultrafast Spectroscopy, Swinburne University of Technology, Melbourne 3122, Australia \\
$^2$Departement de Physique, Ecole Normale Superieure, 24 rue Lhomond, 75005 Paris, France}

\date{\today}

\begin{abstract}

We present an experimental investigation of the dynamic spin response of a strongly interacting Fermi gas using Bragg spectroscopy.  By varying the detuning of the Bragg lasers, we show that it is possible to measure the response in the spin and density channels separately.  At low Bragg energies, the spin response is suppressed due to pairing, whereas the density response is enhanced.  These experiments provide the first independent measurements of the spin-parallel and spin-antiparallel dynamic and static structure factors which provide insight into the different features of density and spin response functions.  At high momentum the spin-antiparallel dynamic structure factor displays a universal high frequency tail, proportional to $\omega^{-5/2}$, where $\hbar \omega$ is the probe energy.  

\end{abstract}


\maketitle

Two-component Fermi gases near Feshbach resonances provide a well controlled setting to explore many-body phenomena in highly correlated quantum systems \cite{Bloch08mbp,Giorgini08tou}. When the interparticle interactions are sufficiently strong, ultracold atomic gases display universal features, where macroscopic parameters become independent of the microscopic details of the interatomic potential \cite{Heiselberg01fsw,Ho04uto,Hu07uto}.  Most studies to date have focussed on static aspects of universality \cite{Tan08eoa,Braaten08erf,Zhang09upo,Hu10ssf,Stewart10vou}; however, certain dynamical properties can also become universal.  Key among these are dynamic susceptibilities which describe the way a system responds to a perturbation.  Recent theoretical work has shown that the dynamic structure factor, which is connected to the imaginary part of susceptibility through the fluctuation-dissipation theorem, shows a universal high frequency tail both at high momentum \cite{Son10sda,Nishida11hpo,Hu12uds} and low momentum where it depends on the frequency dependent shear viscosity \cite{Taylor10vos,Enss11vas}.

Bragg spectroscopy is a well established tool to measure dynamic and static density response functions \cite{Stenger99bso,Brunello01mtt,Steinhauer02eso,Veeravalli08bso}.  Previous work on Fermi gases has shown that the static structure factor follows a universal law which arises from Tan's relation for the density-density correlator \cite{Kuhnle10ubo}.  Several theoretical studies have also investigated the dynamic spin response \cite{Nishida11hpo,Hu12uds,Bruun06bso, Stringari09das,Guo10etp} and a recent study of universal spin transport examined the static spin susceptibility \cite{Sommer11ust,Palestini12das}, yet the dynamic spin susceptibility has not been studied experimentally.  

In this letter we present the first measurements of the dynamic spin response of a strongly interacting Fermi gas.  Two-photon Bragg scattering is used to probe either the spin or density response by appropriate choice of the Bragg laser detuning.  This allows full characterisation of the spin-parallel and spin-antiparallel components of the dynamic and static structure factors through the application of the $f$-sum rule \cite{Pines66tto,Pitaevskii03bec}.  The spin response is suppressed at low energies due to pairing and displays a universal high frequency tail, decaying as $\omega^{-5/2}$, where $\hbar \omega$ is the probe energy (Bragg frequency) \cite{Hu12uds}.

The key to accessing the spin response in two-photon scattering experiments is to use Bragg lasers with a different coupling to each of the two spin states in the mixture.  This can be achieved using spin-flip Bragg spectroscopy \cite{Rodriguez02lic,Bruun06bso}, polarisation sensitive coupling \cite{Carusotto06bsa} or by detuning the Bragg lasers close to resonance.  In our experiments with $^6$Li, the first two methods prove challenging because of the atomic state configuration at high magnetic fields, and therefore we use the third method.  

To understand our measurements we first review the atomic level structure of $^6$Li atoms.  Atoms are cooled to degeneracy in an equal mixture of the two ground states $| F = 1/2, m_F = \pm 1/2 \rangle$ labelled $\mid \uparrow \rangle$ and $\mid \downarrow \rangle$.  At magnetic fields near the broad Feshbach resonance at 834 G the electronic and nuclear spins are almost fully decoupled and atoms undergo transitions which preserve the nuclear spin projection $m_I$.  Both states $\mid \uparrow \rangle$ and $\mid \downarrow \rangle$ have electronic angular momentum $m_J = -1/2$ which can couple to excited states with $m_{J'} = 1/2,-1/2$ and $-3/2$.  At these magnetic fields the splitting, $\omega_{\uparrow \downarrow}$, between states $\mid \uparrow \rangle$ and $\mid \downarrow \rangle$ is approximately 80 MHz and the splitting between excited electronic states is of order 1.5 GHz.  Choosing the polarisation of the Bragg beams so that only $\sigma^-$ transitions are possible ($m_J = -1/2$ to $m_{J'} = -3/2$), combined with the large Zeeman splitting between excited states, it is straightforward to isolate one, effectively closed, atomic transition from each ground state to play any role in the Bragg process.

In the experiments which follow we work in a regime where the Bragg lasers are weak enough to not significantly deplete the cloud.  We also use a long Bragg pulse such that its Fourier width is narrow compared to the spectral features being measured.  In this limit the momentum transferred by the Bragg lasers is proportional to the imaginary part of the dynamic susceptibility, $\chi''(\mathbf{k},\omega) = \pi [S(\mathbf{k},\omega)-S(-\mathbf{k},-\omega)]$ \cite{Brunello01mtt,Stringari09das}, where $S(\mathbf{k},\omega)$ is the dynamic structure factor.  When the magnitude of the Bragg wavevector $| \mathbf{k} |$ is large compared to the Fermi wavevector $k_F$ and the inverse of the de Broglie wavelength $\lambda_{dB}^{-1}$ only the positive term, $S(\mathbf{k},\omega)$, contributes to the measured response.  Thus we can directly obtain $S(\mathbf{k},\omega)$ from a Bragg spectrum.  In three-dimensional Fermi gases $S(\mathbf{k},\omega)$ depends only $| \mathbf{k} | = k$.

The perturbation introduced by the Bragg lasers can be expressed as the sum of two terms, one which couples to the total density $\hat{\rho}(k) = \hat{\rho}_\uparrow(k) + \hat{\rho}_\downarrow(k)$ and another which couples to the total spin $\hat{\mathbf{S}}(k)$.  The strength of the coupling to each spin state is determined by the light shift potential $V_\sigma = | d_{eg} |^2 E^2 / (\hbar \Delta_\sigma)$, where $\Delta_\sigma$ is the detuning from state $\sigma = \;  \uparrow,\downarrow$, $d_{eg}$ is the electric dipole operator for the $m_J = -1/2$ to $m_{J'} = -3/2$ transition and $E$ is the electric field amplitude.  Choosing the $z$-axis parallel to the Feshbach magnetic field the effective Bragg Hamiltonian is given by 
\begin{equation}
H_{\mathrm{eff}} = \left [ \frac{\mathsfsl{I}_{\mathrm{eff}}}{2} \hat{\rho}(k) + \frac{\mathsfsl{B}_{\mathrm{eff}}}{2} \hat{S_z}(k) \right ] e^{-i \omega t} + h.c.,
\end{equation}
where $\mathsfsl{I}_{\mathrm{eff}} = V_{\uparrow} + V_{\downarrow}$ is an effective total intensity given by the sum of the light shifts for each spin state, $\mathsfsl{B}_{\mathrm{eff}} = V_{\uparrow} - V_{\downarrow}$ is an effective magnetic field given by the differential light shift \cite{Cohen72eso}, and $\hat{S_z}(k) = \hat{\rho}_\uparrow(k) - \hat{\rho}_\downarrow(k)$ is the $z$-projection of the total spin.  The response in the density channel is proportional to the total coupling of the Bragg lasers and the spin response is proportional to the difference between the couplings to the different spin states.  As $\omega_{\uparrow \downarrow} \sim 80$ MHz, detunings close to resonance must be employed to access the spin response.  This can lead to spontaneous emission and heating for relatively modest Bragg laser intensities; however, it is still possible to access the linear response regime.

In a spin-balanced two-component Fermi gas the density (spin) dynamic structure factor is defined as  \cite{Combescot06msi}
\begin{equation}
S_{D+,(S-)}(k,\omega) = 2 \left [  S_{\uparrow \uparrow}(k,\omega) \pm S_{\uparrow \downarrow}(k,\omega) \right ],
\end{equation}
and the components of $S(k,\omega)$ are given by
\begin{equation}
S_{\sigma \sigma'}(k,\omega) = \frac{1}{2 \pi N} \int e^{-i \omega t} \langle \hat{\rho}_{\sigma}(k,t)  \hat{\rho}_{\sigma'}^{\dagger}(k,0) \rangle dt,
\end{equation}
where $ \hat{\rho}_{\sigma}(k,t)$ is the time-dependent density operator for state $\sigma$ \cite{Combescot06msi,Griffin93eia}.

For the case of $^6$Li at large magnetic field, both the $\mid \uparrow \rangle$ and $\mid \downarrow \rangle$ states have essentially identical dipole matrix elements $d_{eg}$ for the respective Bragg transitions, and therefore the Bragg laser detunings $\Delta_{\uparrow}$ and $\Delta_{\downarrow}$ determine the relative strength of the density and spin perturbations.  In a spin-balanced Fermi gas $S_{\uparrow \uparrow}(k,\omega) = S_{\downarrow \downarrow}(k,\omega)$ and $S_{\uparrow \downarrow}(k,\omega) = S_{\downarrow \uparrow}(k,\omega)$ \cite{Combescot06msi}, and recalling that $S(-k,-\omega) \rightarrow 0$ for $k \gg k_F, \lambda_{dB}^{-1}$ in the frequency range we measure, the golden rule \cite{Pines66tto} gives the momentum transferred to the atoms by the Bragg perturbation to be proportional to
\begin{equation}
{\cal P}(k,\omega) \propto  \left ( \frac{1}{\Delta_{\uparrow}^2} + \frac{1}{\Delta_{\downarrow}^2} \right ) S_{\uparrow \uparrow}(k,\omega) + \frac{2}{\Delta_{\uparrow} \Delta_{\downarrow}}  S_{\uparrow \downarrow}(k,\omega).
\end{equation}
From Eq. (4) one can see that using a large detuning from resonance, such that $\Delta_{\uparrow} \approx \Delta_{\downarrow} (\gg \omega_{\uparrow \downarrow})$ couples primarily to the density channel as the coefficients in front of $S_{\uparrow \uparrow}$ and $S_{\uparrow \downarrow}$ will be approximately equal.  Conversely, if $\Delta_{\uparrow} = - \Delta_{\downarrow}$ then the response will be entirely in the spin channel.  Choosing detunings between these limits gives a variable contribution of both spin and density response.

The quantities on the right hand side of Eq. (4) each satisfy sum rules given by \cite{Guo10etp}
\begin{gather}
\int \omega S_{\uparrow \uparrow}(k,\omega) d \omega = \frac{N}{\hbar} \omega_r \\ \int \omega S_{\uparrow \downarrow}(k,\omega) d \omega = 0,
\end{gather}
where $N$ is the total number of particles, $\omega_r = \hbar k^2 / 2 m$ is the atomic recoil frequency and $m$ is the atomic mass. Thus, the first energy weighted moment of any measured response must also satisfy the $f$-sum rule which allows the accurate normalisation of our measured Bragg spectra \cite{Kuhnle10ubo}.

In our experiments we prepare ultracold Fermi gases in a balanced mixture of the $\mid \uparrow \rangle$ and $\mid \downarrow \rangle$ states with $N_\sigma = 2 \times 10^5$ atoms per state as described elsewhere \cite{Fuchs07mbe}.  These clouds are loaded into a single beam optical dipole trap with trapping frequencies of 97 and 24.5 Hz in the radial and axial directions, respectively.  Bragg scattering is achieved by illuminating the cloud with two beams intersecting at an angle of $84^\circ$ giving $\omega_r/(2 \pi) = 132$ kHz and a relative Bragg wavevector of $k/k_F = 4.5$.  We use Bragg laser detunings of $\Delta_\uparrow \, (\Delta_\downarrow) = 584 \, (662) $ MHz to measure the density response and $\Delta_\uparrow \, (\Delta_\downarrow) = -(+) \,$39 MHz for the spin.  These detunings, particularly in the spin case, are relatively small and can lead to heating via spontaneous emission.  For significant intensities this could alter the density and spin correlations that determine the response.  To establish when this occurs we measure the Bragg response for a range of Bragg frequencies in both the spin and density channels as a function of the Bragg laser intensity as described in the supplemental material.  Knowing the regimes of linear response, it is possible to measure the linear density and spin response functions using optimised Bragg laser intensities.  

\begin{figure}[htbp]
\centering
\includegraphics[clip=true,width=0.47\textwidth]{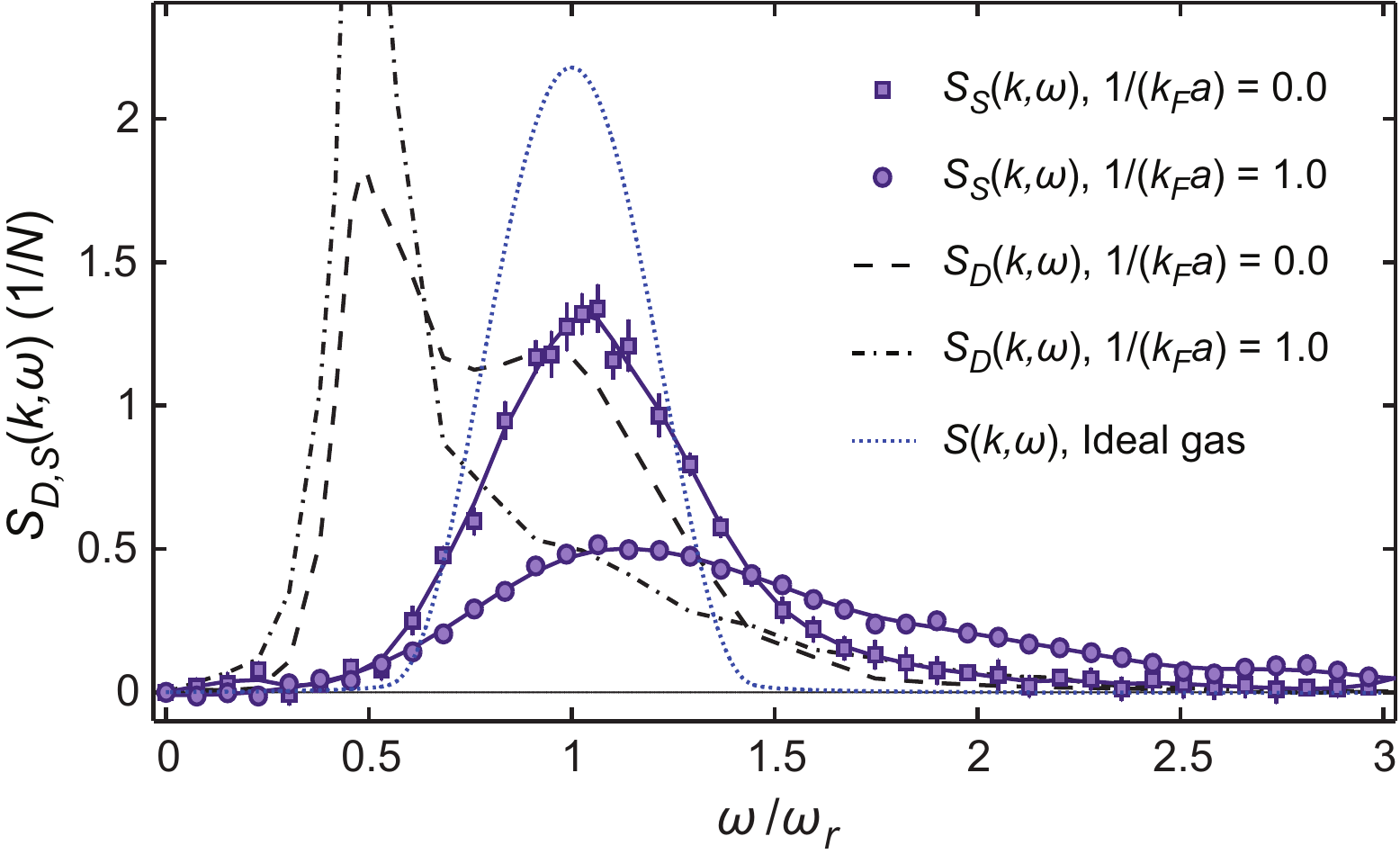}
\caption{Dynamic spin and density structure factors of a two-component Fermi gas.  The spin (filled squares and circles) and density response (dashed and dot dashed lines) where measured at $1/(k_F a) = 0.0$ and $1.0$, respectively, and the dotted blue line is the expected response of a noninteracting Fermi gas at $T = 0$.  Solid lines are smoothed curves through the spin response data to guide the eye and the responses have been normalised using the $f$-sum rule.  }
\label{linearity}
\end{figure}

Spectra are obtained by applying a short Bragg pulse to a trapped atom cloud and measuring the momentum imparted as a function of the Bragg frequency.   Following the Bragg pulse the trap is immediately turned off and the atoms are allowed to expand for 500 $\mu$s before an absorption image of atoms in state $\mid \uparrow \rangle$ is taken.  The imaging laser frequency is then rapidly switched and a second image of atoms in state $\mid \downarrow \rangle$ is taken 850 $\mu$s after the first.  Taking separate images of each spin state at different times allows us to measure the differential center of mass cloud displacement which is insensitive to fluctuations in the trap position.  As we use a balanced mixture, and with the large collisional coupling between the two spin states, the order in which we image states $\mid \uparrow \rangle$ and $\mid \downarrow \rangle$ makes no difference to the measured spectra.

Figure 1 shows Bragg spectra for both the spin (filled squares and circles) and density (dashed and dash-dotted lines) at unitarity, $1/(k_F a) = 0.0$, where $a$ is the $s$-wave scattering length, and on the BEC side, $1/(k_F a) = 1.0$.  Also shown for comparison is the Bragg response of an ideal Fermi gas at $T = 0$ (blue dotted line).  These spectra were obtained at the lowest temperatures achievable with our experiment ($\sim 0.06 \, T_F$ at unitarity).  Each spectrum is normalised such that the first energy weighted moment, $\int \omega S(k,\omega) d \omega$, is set to be equal to unity.  According to the $f$-sum rule  \cite{Pines66tto}, this is equivalent to dividing each spectrum by the total number of particles multiplied by the recoil energy.  Therefore by normalising in this way, and expressing $\omega$ in units of $\omega_r$, we obtain the dynamic structure factor in units of $1/N$ \cite{Kuhnle10ubo}. 

Both the spin and density responses are significantly different to the ideal gas case indicating the strong correlations present both in the BEC and unitarity regimes.  The density spectra $S_D(k,\omega)$ show a strong response at $\omega_r / 2$ due to the scattering of spin-up/spin-down particle pairs, as well as, a broad response at higher frequencies in the region of $\omega_r$ due to the scattering of single atoms \cite{Veeravalli08bso}.  At high momentum the collective mode evolves into the pairing feature at $\omega_r/2$ \cite{Combescot06msi}, which is clear in both spectra, but more prominent on the BEC side due to the increased likelihood of finding spin-up/spin-down particles at small separation.  

Spin Bragg spectroscopy, on the other hand, is not sensitive to the collective (paired) mode as the positive and negative perturbations on the spin-up and spin-down particle, respectively, cancel each other, leaving no nett perturbation on a pair.  Therefore the spin response $S_S(k,\omega)$ is suppressed at $\omega_r/2$ and instead only shows free atom excitations at higher energies.  $S_S(k,\omega)$ on the BEC side has a lower peak and is biased towards higher frequencies showing a suppression of the spin susceptibility due to the increased energy required to remove atoms from bound pairs.  

\begin{figure}[htbp]
\centering
\includegraphics[clip=true,width=0.47\textwidth]{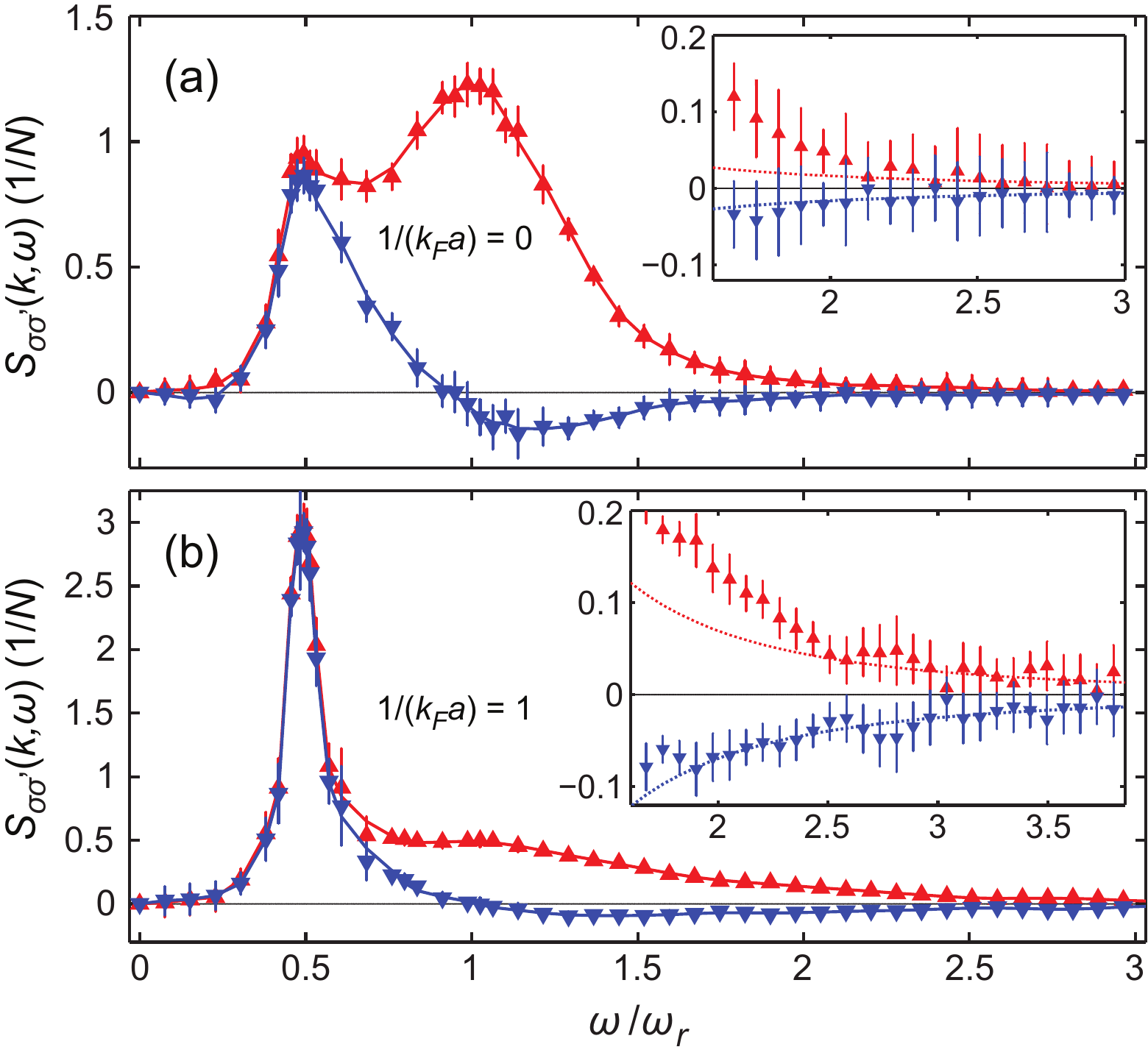}
\caption{Spin-parallel and spin-antiparallel components of the dynamic structure factor of a strongly interacting Fermi gas measured at (a) $1/(k_F a) = 0.0$, and, (b) at $1/(k_F a) = 1.0$.   Red upright triangles are the spin-parallel structure factor $S_{\uparrow \uparrow}$ and blue inverted triangles are the spin anti-parallel response $S_{\uparrow \downarrow}$.  Solid lines are a guide to the eye.  Insets show zoomed in plots of the high frequency region where $S_{\uparrow \downarrow}(k,\omega)$ shows a universal tail proportional to $\omega^{-5/2}$.  The dotted blue lines are power law fits to $S_{\uparrow \downarrow}(k,\omega > 2\omega_r)$ and the dotted red line is the negative of these fits, indicating that $S_{\uparrow \uparrow}(k,\omega)$ only approaches the asymptotic behaviour at higher $\omega$.}
\label{spectra}
\end{figure}

The normalised spectra in Fig. 1 can be combined with the known Bragg laser detunings according to Eq. (4) to give the spin-parallel $\{ (S_{\uparrow \uparrow}(k,\omega)$, red triangles$\}$ and spin-antiparallel $\{ (S_{\uparrow \downarrow}(k,\omega)$, blue circles$\}$ components of the dynamic structure factor, plotted in Fig. 2(a) and (b) at $1/(k_F a) = 0.0$ and 1.0, respectively.  These spectra reveal the response of the particle density in one state to a perturbation of the particle density in either the same or opposite spin states and show quite a complex structure.  We note that in the absence of interactions, the spin-parallel response would be identical to the ideal gas response shown in Fig. 1 and the spin-antiparallel response would be zero for all $\omega$.

The features seen in $S_{\uparrow \uparrow}(k,\omega)$ and $S_{\uparrow \downarrow}(k,\omega)$ can now be examined and attributed to different excitations.  At frequencies ranging from $0$ to slightly above $\omega_r/2$, both the spin-parallel and spin-antiparallel components are essentially identical for both interaction strengths.  This striking result shows that in this frequency range the measured response is entirely dominated by the scattering of spin-up/spin-down pairs.  Each pair contains a  spin-up and spin-down particle, and, since the response is independent of the combination of spin states being probed, the density-density correlations for either combination of spin states are the same.

At higher frequencies, however, the spin-parallel and spin-antiparallel response functions become very different with $S_{\uparrow \uparrow}(k,\omega)$ showing a significant peak near $\omega_r$ while  $S_{\uparrow \downarrow}(k,\omega)$ becomes negative for $\omega \gtrsim \omega_r$.  The positive peak in $S_{\uparrow \uparrow}(k,\omega \approx \omega_r)$ is primarily due to the autocorrelation term, that arises from single particle scattering.  This feature is not present in $S_{\uparrow \downarrow}(k,\omega)$ that always involves correlations between different particles.  $S_{\uparrow \downarrow}(k,\omega)$ becomes negative at high frequencies as required to satisfy the sum rule, Eq. (6) which reveals an interesting feature of the relative dynamics of spin-up/spin-down particles.  A high frequency fluctuation with momentum $k$ of the spin-down density leads to an out-of-phase fluctuation of the spin-up particle density.  These anti-correlations suppress the high frequency response in the density channel but enhance the high frequency spin response.

For frequencies much greater than $\omega_r$ $S_{\uparrow \uparrow}(k,\omega)$ $\{S_{\uparrow \downarrow}(k,\omega) \}$ is predicted to display a universal tail proportional to $+\{-\} \, {\cal I} / \omega^{5/2}$, where $\cal I$ is Tan's contact parameter \cite{Hu12uds}.  When measuring the density response alone, these two dependencies cancel leaving a universal $\omega^{-7/2}$ high frequency tail in $S_{D}(k,\omega)$ \cite{Son10sda} reminiscent of neutron scattering measurements on superfluid $^4$He \cite{Griffin93eia}.  The insets of Fig. 2(a) and (b) show zoomed in plots of $S_{\uparrow \uparrow}(k,\omega)$ and $S_{\uparrow \downarrow}(k,\omega)$ at high frequency.  Despite the small signal we find that the tail of $S_{\uparrow \downarrow}(k,\omega)$  is well described by an $\omega^{-5/2}$ dependence.  A free power-law fit to the data (dotted blue line) for $\omega > 2 \omega_r$ yields $S_{\uparrow \downarrow}(k,\omega) = (-0.39 \pm 0.11) \,  \omega^{-2.5 \pm 0.3}$ on the BEC side and $S_{\uparrow \downarrow}(k,\omega) = (-0.14  \pm 0.10)  \, \omega^{-2.7 \pm 0.8}$ at unitarity, consistent with $\omega^{-5/2}$.  

In the limit that $k,\omega \rightarrow \infty$, the amplitude of the tails should provide a measure of the contact; however, our fit coefficients are approximately a factor of three larger than expected \cite{Son10sda,Hu12uds}.  This can be partially attributed to the fact that the measured response is given by the convolution of the dynamic structure factor with a squared sinc function \cite{Veeravalli08bso,Brunello01mtt} which will lift the data; however, we estimate that this effect alone would not explain the discrepancy.  It is more likely that our experiments are not in the strict $k,\omega \rightarrow \infty$ limit to obtain the contact even though the power law describes the data well.  We note that $S_{\uparrow \uparrow}(k,\omega)$ only approaches the asymptotic behaviour at higher Bragg frequencies (dotted red line) so we do not perform a separate fit to $S_{\uparrow \uparrow}(k,\omega)$, nor do we resolve the $\omega^{-7/2}$ tail in $S_{D}(k,\omega)$ in our measured frequency range \cite{Son10sda}.

Finally, the spectra in Fig. 2 can be integrated to provide the static structure factors $S(k)$.  As these spectra are expressed in units of $1/N$, integration over $\omega$ gives $S(k)$ directly \cite{Kuhnle10ubo}.  We find $S_{\uparrow \uparrow}(k) \: \{ S_{\uparrow \downarrow}(k) \}$ to be $1.02 \pm 0.04 \: \{0.17 \pm 0.04 \}$ at unitarity and  $1.03 \pm 0.04 \: \{0.38 \pm 0.04 \}$ at $1/(k_F a) = 1.0$.  At the momentum used here $S_{\uparrow \uparrow}(k)$ is expected to be unity \cite{Combescot06msi} consistent with our findings.  This also indicates that both the spin and density responses provide consistent measures for $S_{\uparrow \downarrow}(k)$ within our uncertainties.  The uncertainties in the combined spectra are dominated by the uncertainty in the spin response, which was obtained at low intensities to minimise spontaneous emission.  The situation could be improved by using a spin mixture with a larger energy separation such as $| F = 1/2, m_F = 1/2 \rangle$ and $| F = 3/2, m_F = - 3/2 \rangle$ \cite{Bartenstein05pdo}.

In summary we have shown that Bragg spectroscopy with a differential coupling to each spin state can measure the dynamic spin and density responses of a strongly interacting two-copmonent Fermi gas.  We have used this technique to extract $S_{\uparrow \uparrow}(k,\omega)$ and $S_{\uparrow \downarrow}(k,\omega)$ separately which allows us to associate the different features of the density and spin response with the scattering of pairs and free atoms.  The spin-antiparallel dynamic structure factor was seen to display a universal high frequency tail proportional to $\omega^{-5/2}$.  This work opens the way to measurements of $S_{\uparrow \uparrow}(k,\omega)$ and $S_{\uparrow \downarrow}(k,\omega)$ at any momentum, which will be particularly important for $k \lesssim k_F$ where $S_{\uparrow \uparrow}(k) < 1$.
 
We thank H. Hu and P. Hannaford for stimulating discussions.  This work is supported by the Australian Research Council.

\end{document}